\documentclass[aps,superscriptaddress,showpacs,nofootinbib,floatfix,epfs]{revtex4}
\pdfoutput=1
\usepackage{amsfonts,amscd,amsmath, amssymb,graphicx,color}

\newcommand{\oh}{\frac{1}{2}}
\def\g3{\mathbf{g}}
\def\G3{\boldsymbol{\Gamma}}
\def\oh{\frac{1}{2}}
\def\A{\cal{A}}
\begin{document}
\title{On Nonrelativistic Diffeomorphism Invariance}
\author{Oleg Andreev}
\affiliation{Arnold Sommerfeld Center for Theoretical Physics, LMU-M\"unchen, Theresienstrasse 37, 80333 M\"unchen, Germany}
\affiliation{L.D. Landau Institute for Theoretical Physics, Kosygina 2, 119334 Moscow, Russia}
\author{Michael Haack}
\affiliation{Arnold Sommerfeld Center for Theoretical Physics, LMU-M\"unchen, Theresienstrasse 37, 80333 M\"unchen, Germany}
\author{Stefan Hofmann}
\affiliation{Arnold Sommerfeld Center for Theoretical Physics, LMU-M\"unchen, Theresienstrasse 37, 80333 M\"unchen, Germany}
%\date{\empty}
\begin{abstract} 
We study certain aspects of the recently proposed notion of nonrelativistic diffeomorphism invariance. In particular, we consider
specific examples of invariant actions, extended gauge symmetry as well as an application to the theory of quantum Hall effect. We also discuss an alternative approach based on Ho\v{r}ava-Lifshitz gravity. 
\end{abstract}
%\preprint{LMU-ASC 78/12}
\maketitle

\vspace{-7.5cm}
\begin{flushright}
LMU-ASC 68/13
\end{flushright}
\vspace{6cm}

\vspace{.5cm}
%____________________________________________________ section 1
\section{Introduction}
\label{intro}
\renewcommand{\theequation}{1.\arabic{equation}}
\setcounter{equation}{0}

The method of effective field theory is a powerful tool. One can use it to describe physics at a given energy scale to a given accuracy in terms of a quantum field theory with a finite set of parameters. Moreover, one can formulate the effective field theory without any reference to what goes on at arbitrarily small distances. In practice, one writes down the most general possible Lagrangian including all terms, renormalizable or not, consistent with assumed symmetry principles. From this point of view, the existence of symmetry is of great value in constructing explicit examples of effective field theories. The symmetry might lead to non-trivial constraints on effective Lagrangians (especially for higher order terms) and, as a consequence, some of the free parameters would be fixed without matching predictions of the effective theories to those of the true theories.

An important example to which this discussion applies is a nonrelativistic theory living on a curved $d$-dimensional manifold with the metric $g_{ij}=g_{ij}(t,\mathbf{x})$ in the presence of a $U(1)$ gauge field $A_\mu =(A_0(t,\mathbf{x}),\,A_i(t,\mathbf{x}))$. Here $\mathbf{x}=x^i$ and $i=1,\dots , d$. A remarkably simple suggestion made in \cite{SW} is that under diffeomorphisms generated by 
\begin{equation}\label{xi}
\delta t=0\,,\qquad\delta x^i=\xi^i(t,\mathbf{x})
\end{equation}
the background fields transform as\footnote{To get the form as written in \cite{SW}, one has to replace $m\rightarrow -m$.}
\begin{equation}\label{transNR}
\delta A_0=-\partial_k A_0\xi^k- A_k\dot\xi^k\,,\qquad\delta A_i=-\partial_k A_i\xi^k- A_k\partial_i\xi^k-mg_{ik}\dot\xi^k\,,\qquad
\delta g_{ij}=-\partial_k g_{ij}\xi^k-g_{kj}\partial_i\xi^k-g_{ik}\partial_j\xi^k\,,
\end{equation}
where $\dot{\xi}=\partial_t\xi$ and $m$ is a mass parameter. In addition, as usual, the theory is supposed to be invariant under $U(1)$ gauge transformations generated by 
\begin{equation}\label{transG}
\delta A_0=-\dot{\alpha}\,,\qquad
\delta A_i=-\partial_i\alpha\,,\qquad
\delta g_{ij}=0\,.
\end{equation}

The discussion in \cite{SW} was motivated by consideration of a system of noninteracting (spin-zero) particles. In the presence of the background fields, the corresponding action is given by
\begin{equation}\label{Schroedinger}
S=\frac12 \int dtd\mathbf{x}\sqrt{g}\biggl[i \psi^\dagger\overset\leftrightarrow{\partial}_t\psi+2 A_0\psi^\dagger\psi-
\frac{g^{ij}}{m}\bigl(\partial_i\psi^\dagger+iA_i\psi^\dagger\bigr)
\bigl(\partial_j\psi-iA_j\psi\bigr)
\biggr]	
\,,
\end{equation}
where $d\mathbf{x}=dx^1\dots dx^d$ and $\psi^\dagger\overset\leftrightarrow{\partial}_t\psi=\psi^\dagger\dot\psi-\dot\psi^\dagger\psi$. A simple analysis shows that \eqref{Schroedinger} is invariant under \eqref{transNR} and \eqref{transG} accompanied by the transformation of $\psi$ 
\begin{equation}\label{psi}
\delta\psi=-i\alpha\psi-\partial_k\psi\xi^k \,.
\end{equation}

Alternatively, the transformation rules \eqref{transNR} and \eqref{transG} appear \cite{SW} as the leading order in the $1/c$ expansion of relativistic diffeomorphisms for the spacetime metric $g_{\mu\nu}$ 
\begin{equation*}
\delta g_{\mu\nu}=-\partial_\lambda g_{\mu\nu}\xi^\lambda-g_{\lambda\nu}\partial_\mu\xi^\lambda-g_{\mu\lambda}\partial_\nu\xi^\lambda
\,,
\end{equation*}
where $\xi^\lambda=\bigl(-\frac{\alpha}{mc},\xi^k\bigr)$, $x^0=ct$ and ${\mu,\nu}=0,\dots,d$. For such an expansion, the ansatz for the 
metric is given by 
\begin{equation}\label{metricd+1}
g_{\mu\nu}=
\begin{pmatrix}
-1+{\displaystyle\frac{2A_0}{mc^2}}\,\,&{}\,\,&{\displaystyle\frac{A_i}{mc}}\,\,\\
{}\,\,&{}\,\,&\,\,\\
{\displaystyle\frac{A_i}{mc}}\,\,&{}\,\,& g_{ij}\,\,
\end{pmatrix}\,,
\end{equation}
with $A_0$ and $A_i$ kept fixed as $c\rightarrow\infty$.

How can one include the above symmetries in the formulation of the effective field theory? This can be done quite simply by integrating out the dynamical variables and then considering the resulting effective action as a functional of the background fields. Thus, there is no need for 
the explicit transformation rules of the dynamical fields. An important example to which this idea applies is the Hall liquid \cite{HS}. In this case, it allows one to relate the Hall viscosity with the leading correction to the Hall conductivity (in an expansion in small wave numbers). 

In the present paper, we will study certain aspects of the nonrelativistic diffeomorphism invariance. For orientation, we begin by describing the transformation rules \eqref{transNR} from the viewpoint of foliation-preserving diffeomorphisms. In section III, we give some examples of invariant actions. We go on in Section IV to discuss gauge symmetry enhancement in a theory with fermions. In section V, we consider some applications to the Hall fluid. In particular, we show how the continuity equation follows from the invariance of an effective action under nonrelativistic diffeomorphisms. The continuity equation in turn implies a relation between the components of the conductivity tensor and other transport coefficients \cite{Read}. In section VI, we consider some related issues in Ho\v{r}ava-Lifshitz gravity and propose an alternative way of constructing invariant actions. We conclude in section VII with a discussion of some open problems and possibilities for further study. Some technical details are given in the appendix.

%__________________________________________________________ section 2
\section{Preliminaries} 
\renewcommand{\theequation}{2.\arabic{equation}}
\setcounter{equation}{0}

The transformations \eqref{xi} are a subgroup of foliation-preserving diffeomorphisms\footnote{In section VI, we will discuss these issues in light of Ho\v{r}ava-Lifshitz gravity.}, with the time coordinate $x^0$ chosen to be a global time. In this case, the spacetime metric takes the form
\begin{equation}\label{adm}
	ds^2=-c^2dt^2+g_{ij}\bigl(dx^i-V^idt\bigr)\bigl(dx^j-V^j dt\bigr)
	\,.
\end{equation}
Once one specifies the spacetime metric in terms of the spatial metric $g_{ij}$ and the shift vector $V_i$, one can easily find the transformation rules
\begin{equation}\label{transV}
\delta V^i=-\partial_k V^i\xi^k+V^k\partial_k\xi^i+\dot\xi^i\,,\qquad
\delta g_{ij}=-\partial_k g_{ij}\xi^k-g_{kj}\partial_i\xi^k-g_{ik}\partial_j\xi^k\,.
\end{equation}

In this geometry, the action for a nonrelativistic system of noninteracting particles takes the form
\begin{equation}\label{Schr-inv}
S=\frac12 \int dtd\mathbf{x}\sqrt{g}\biggl[i\psi^\dagger\overset\leftrightarrow{\partial}_t\psi+mV_iV^i\psi^\dagger\psi-
\frac{g^{ij}}{m}\bigl(\partial_i\psi^\dagger-imV_i\psi^\dagger\bigr)
\bigl(\partial_j\psi+imV_j\psi\bigr)
\biggr]	
\,,
\end{equation}
where $V_i=g_{ij}V^j$. It is invariant under \eqref{transV}, with $\delta\psi$ having the form of eq.\eqref{psi}.
Given the action that we have just described, it is straightforward to include an external electromagnetic field $A_\mu^{\text{\tiny elm}}$ which transforms under nonrelativistic diffeomorphisms in a standard way (as a 1-form). This is done by replacing the derivatives: $\partial_t\rightarrow\partial_t-iA_0^{\text{\tiny elm}}$ and $\partial_i\rightarrow\partial_i-iA_i^{\text{\tiny elm}}$. Now we see that if the electromagnetic field is combined with the shift vector to define a field $A_\mu$ 
\begin{equation}\label{gauge-field}
	A_0=A_0^{\text{\tiny elm}}+\frac{1}{2}mV_iV^i\,,\qquad
	A_i=A_i^{\text{\tiny elm}}-mV_i\,,
\end{equation}
then the action \eqref{Schr-inv} reduces to that of eq.\ \eqref{Schroedinger}. Moreover, such a field transforms according to the transformation rules \eqref{transNR} and \eqref{transG}, as expected. This is similar to the discussion of \cite{Son,Sumrule}, where the Hall fluid living in Newton-Cartan geometry is discussed.

In the present paper, we study gauge invariant actions in terms of the gauge field $A_\mu$ and the spatial metric $g_{ij}$. Obviously, 
requiring invariance under nonrelativistic diffeomorphims imposes constraints on the actions. At the end of the calculations, one could set $V^i=0$ in order to obtain the actions in terms of the ordinary background electromagnetic field and the spatial metric.

Note that Galilei transformations are a subset of \eqref{xi}. In quantum mechanics the invariance of the Schr\"odinger equation under Galilei transformations is usually implemented via allowing $\psi$ to transform projectively, cf.\ e.g.\ \cite{Ballentine}. However, here we allow the background metric (or more precisely the shift vector $V^i$) to transform, as \eqref{xi} is a subgroup of the diffeomorphism group. As a result,  the Schr\"odinger equation is invariant without the need for a projective transformation of $\psi$.

%__________________________________________________________ section 3
\section{Some Examples of Invariant Actions} 
\renewcommand{\theequation}{3.\arabic{equation}}
\setcounter{equation}{0}

We will now carry out some sample calculations to illustrate the above ideas. We specialize to the case of two spatial dimensions, 
which is the easiest and also of interest for potential applications to condensed matter physics. 

\subsection{The $c\rightarrow\infty$ limit}

In order to find examples of invariant actions, we start with the Einstein-Hilbert action,
\begin{equation}\label{EH3}
	S \sim mc\int d^3x \sqrt{g^{(3)}}\,R^{(3)}\,,
\end{equation}
where $g^{(3)}=\det g_{\mu\nu}$ and $R^{(3)}$ is the corresponding scalar curvature.\footnote{Here and below, we follow the conventions listed in the appendix.} On dimensional grounds, we take $G_N\sim 1/mc$. This action is invariant under the full diffeomorphism group. Using the expansion \eqref{metricd+1} and dropping the $1/c$ terms, we obtain
\begin{equation}\label{EH3a}
	S=4\pi mc^2\int dt\,\chi+ m\int dt d\mathbf{x} \sqrt{g}
	\biggl[-\frac{1}{4}\dot{g}^{ij}\dot{g}_{ij}-\frac{1}{4}(g^{ij}\dot{g}_{ij})^2-
	\frac{1}{m}\Bigl(A^i\partial_t\nabla_i\ln g-\dot{g}^{ij}\nabla_i A_j+A_0 R\Bigr)+
	\frac{1}{4m^2}F_{ij}F^{ij}\biggr]\,.
\end{equation}
In this case, a nonrelativistic theory lives on a 2-dimensional surface without boundaries, endowed with a metric tensor $g_{ij}$ and parameterized by local coordinates $x^i$. So, $g=\det g_{ij}$, $R$ is the corresponding scalar curvature, and $F_{ij}=\partial_iA_j-\partial_j A_i$. $\chi$ denotes the Euler characteristic of the surface. After the subtraction of the first term, which is manifestly invariant under 
the transformations \eqref{transNR} and \eqref{transG}, the action is finite in the limit $c\rightarrow\infty$. 

Now, using the standard formulae for the spin connection in terms of zweibeins together with a definition of $\omega_0$
\begin{equation}\label{SC}
	\omega_0=\frac{1}{2}\varepsilon^{ab}e^j_a\dot{e}^b_j\,,\quad
	\omega_i=\frac{1}{2}\varepsilon^{ab}e^j_a\nabla_i e^b_j\,,
\end{equation}
one can express it (up to a constant multiple) as 
\begin{equation}\label{EH3b}
	S=\int dt d\mathbf{x} \sqrt{g}
	\biggl[\frac{m}{8}\bigl(\dot{g}^{ij}\dot{g}_{ij}+(g^{ij}\dot{g}_{ij})^2\bigr)+
	\frac{1}{\sqrt{g}}\varepsilon^{\mu\nu\lambda}\omega_\mu\partial_\nu A_\lambda-\frac{1}{4m}B^2 
	\biggr]\,,
\end{equation}
where $\varepsilon^{\mu\nu\lambda}$ is the completely antisymmetric symbol. $B$ denotes the magnetic field 
$B=\text{e}^{ij}\partial_i A_j$, with $\text{e}^{ij}$ the completely antisymmetric tensor.

This is our first example of an action which is invariant under both the nonrelativistic diffeomorphisms \eqref{transNR} and 
the gauge transformations \eqref{transG}.\footnote{Note that it is also invariant under the local $SO(2)$-transformations \eqref{Loretz}. We will return to this point in sec.\ \ref{sec_horava_lifshitz}.} Of course, one can explicitly check that \eqref{EH3b} is invariant. An interesting feature is that the Wen-Zee 
term $\omega dA$ \cite{WZ} occurs in the resulting expression. It is well known that such a term plays an important role in the description of topological properties of quantum Hall states in $2+1$ dimensions. In section \ref{continuityequation}, we will also discuss some implications for the quantum Hall effect.\footnote{For a comprehensive review, see \cite{Fradkin} and references therein.}

Given an invariant action, one can construct another invariant action by noting that a gauge field $A'_\mu$ defined by
\begin{equation}\label{newA}
A'_0=A_0+\omega_0-\frac{1}{2m}B\,,\qquad
A'_i=A_i+\omega_i\,
\end{equation}
transforms under the transformations \eqref{transNR} and \eqref{transG} in the same way as $A_\mu$, as follows from \eqref{spinconNR} and \eqref{Omegab}.

For instance, substituting the expressions \eqref{newA} into the action \eqref{EH3b} leads to 
\begin{equation}\label{EH3c}
	S=\int dt d\mathbf{x} \sqrt{g}
	\biggl[\frac{m}{8}\bigl(\dot{g}^{ij}\dot{g}_{ij}+(g^{ij}\dot{g}_{ij})^2\bigr)+
	\frac{1}{\sqrt{g}}\Bigl(\varepsilon^{\mu\nu\lambda}\omega_\mu\partial_\nu A_\lambda+\varepsilon^{\mu\nu\lambda}\omega_\mu\partial_\nu \omega_\lambda\Bigr)
	-\frac{1}{4m}\Bigl(B^2+2RB+\frac{1}{4}R^2\Bigr)\biggr]\,.
\end{equation}
Combining \eqref{EH3b} and \eqref{EH3c}, we finally find that the action 
\begin{equation}\label{GSC}
	S=\int dt d\mathbf{x} \Bigl[\varepsilon^{\mu\nu\lambda}\omega_\mu\partial_\nu\omega_\lambda-
	\frac{1}{2m} \sqrt{g}R B
	\Bigr]\,
\end{equation}
is invariant. Of course, one can verify this statement by using the transformation rules. 

Along this line of thought, the appearance of the gravitational 
Chern-Simons term $\omega d\omega$ seems not surprising. It is related to the Wen-Zee term in the action we started with. Alternatively, one can 
derive the action \eqref{GSC} by taking the limit $c\rightarrow\infty$ of the gravitational Chern-Simons action. 

%__________________________________________________________________
\subsection{The $(\omega,q)\rightarrow (0,0)$ limit}

Now we will discuss the limit of small frequencies and wave numbers. We consider the leading correction to an action invariant under 
time-independent diffeomorphisms.

First, let us remind that there is a good approximation for describing the drift velocity of the Hall fluid, i.e.
\begin{equation}\label{velocity}
	v^i=\text{e}^{ki}\frac{E_k}{B}
	\,.
\end{equation}
Its meaning is simple: if the electric field is small enough, there is a local frame moving at a velocity $v^i$ such that in this frame the 
electric field is locally absent. Under the nonrelativistic diffeomorphisms, the velocity transforms as 
\begin{equation}\label{velocitya}
\delta v^i=-\partial_k v^i\xi^k+v^k\partial_k\xi^i+\dot{\xi}^i+\frac{m}{B}\text{e}^{ij}\partial_t(g_{jk}\dot{\xi}^k)
+\frac{m}{B}v^i\text{e}^{nj}\partial_n(g_{jk}\dot{\xi}^k)
\,.
\end{equation}

Now, let $S_0$ be a gauge invariant action. We will assume that it is also invariant under time-independent diffeomorphisms. In 
other words, its variation under the diffeomorphims \eqref{xi} is $\delta S_0=\int dtd\mathbf{x}\, \tilde{{\cal L}}_i\dot{\xi}^i$. In general $\tilde{{\cal L}}_i$ contains terms with different numbers of derivatives. In what follows we only keep the leading term ${\cal L}_i$, as  we are interested in a derivative expansion.

In the case of ${\cal L}_i$ independent of the electric field $E_i$, our ansatz for an invariant action to lowest order in $\omega$ and $q$ is 
\begin{equation}\label{ansatz}
	S=S_0-\int dtd\mathbf{x}\,{\cal L}_i v^i
\,\,.
\end{equation}
As a check, we find that the variation of $\eqref{ansatz}$ becomes zero after dropping all higher order terms in $\omega$ and $q$. In doing so, we assume that $v^i$ is of the same order as $\dot{\xi}^i$.

Likewise, for ${\cal L}_i$ linear in $E_i$, our ansatz is
\begin{equation}\label{ansatzv}
	S=S_0-\oh\int dtd\mathbf{x}\,{\cal L}_{ij} v^iv^j
\,\,,
\end{equation}
with ${\cal L}_i={\cal L}_{ij}v^j$. Also \eqref{ansatzv} is invariant up to higher order terms.

To illustrate the above ideas, we consider a few examples of $S_0$. Using our ans\"atze, we get
\begin{equation}\label{Son-1}
S=\int dtd\mathbf{x}\Bigl[\varepsilon^{\mu\nu\lambda}\omega_\mu\partial_\nu A_\lambda-\oh\varepsilon^{ij}\partial_iB v_j\Bigr]+\dots
\,,\qquad
S=\int dtd\mathbf{x}\Bigl[\sqrt{g}\epsilon(B)-m\epsilon''(B)\varepsilon^{ij}\partial_iB v_j\Bigr]+\dots
\,,
\end{equation}
and 
\begin{equation}\label{Son-2}
S=\int dtd\mathbf{x}\Bigl[\varepsilon^{\mu\nu\lambda}A_\mu\partial_\nu A_\lambda+ m\sqrt{g}B g_{ij}v^iv^j\Bigr]+\dots\,
	\,.
\end{equation}
Here a prime denotes a derivative with respect to $B$. $\epsilon (B)$ is a function of $B$.

We close this section with a few short comments:

\noindent (i) In \eqref{Son-1} one can reproduce the corrections to the $S_0$'s by making a transformation $\xi^i=-v^i t$. To leading order in derivatives, this can be regarded as a Galilean transformation, as is obvious from $\dot{\xi}^i=-v^i+\partial_t(\text{e}^{ik}E_k/B)t$. 

\noindent (ii) The expressions \eqref{Son-1} and \eqref{Son-2} have already been considered in \cite{HS} in the 
context of the quantum Hall effect. We will return to this issue in section \ref{continuityequation}.

\noindent (iii) Interestingly, with the introduction of the vorticity $\Omega=\text{e}^{ij}\partial_i v_j$, one can put \eqref{Son-1} in the form
\begin{equation}\label{Son-1a}
S=\int dtd\mathbf{x}\Bigl[\varepsilon^{\mu\nu\lambda}\omega_\mu\partial_\nu A_\lambda
+\oh\sqrt{g}B\Omega\Bigr]+\dots
\,,\qquad
S=\int dtd\mathbf{x}\sqrt{g}\Bigl[\epsilon(B)+m\epsilon'(B)\Omega\Bigr]+\dots\,
\,.
\end{equation}
To linear order in $\Omega$, the last expression can be rewritten as $S=\int dtd\mathbf{x}\sqrt{g}\epsilon(B+m\Omega)$. This makes it clear that our construction is valid as long as the vorticity is smaller than the cyclotron frequency $\omega_c=B/m$.  

%__________________________________________________________ section 4
\section{Extended gauge symmetry} 
\label{gaugeenhancement}
\renewcommand{\theequation}{4.\arabic{equation}}
\setcounter{equation}{0}

So far in our discussion, we have assumed a single gauge field. However, it can be extended to the case of multiple gauge fields which we would like to do in this section.

For orientation, we consider the generalization of the action \eqref{Schroedinger} to a spinor field. The corresponding action is given by 
\begin{equation}\label{Pauli}
S=\int dtd\mathbf{x}\sqrt{g}\biggl[\frac{i}{2}\Psi^\dagger\overset\leftrightarrow{\partial}_t\Psi+
\Psi^\dagger(A_0+i{\cal A}_0)\Psi-
\frac{g^{ij}}{2m}\bigl(\partial_i\Psi^\dagger+i\Psi^\dagger(A_i+i{\cal A}_i)\bigr)
\bigl(\partial_j\Psi-i(A_j+i{\cal A}_j)\Psi\bigr)
\biggr]
\,,
\end{equation}
where $\Psi$ is the spinor and ${\cal A}_\mu$ is the $SO(d)$ gauge field. 

It is easy to see that the action is gauge invariant under the local $SO(d)\times U(1)$ gauge symmetry group. Indeed, in addition to \eqref{transG}, the rules for the local $SO(d)$ transformation are given by 
\begin{equation}\label{transGa}
\delta {\cal A}_0=-\dot{U}-[{\cal A}_0,U]
\,,\qquad
\delta{\cal A}_i=-\partial_i U-[{\cal A}_i,U]
\,,\qquad \delta g_{ij}=0\,,
\qquad \delta A_\mu=0
\,,
\end{equation}
with $\delta\Psi=U\Psi$.

There is a subtle issue here. Certainly, ${\cal A}_\mu$ is related to a spin connection. If a $d$-dimensional manifold has torsion, then ${\cal A}$ 
may be regarded as an independent gauge field and the settings of section I are needed to be customized to include a new geometric structure. In this case, the effective action is a functional of $A_\mu$, $g_{ij}$, and ${\cal A}_\mu$. If a $d$-dimensional manifold is torsion free, then ${\cal A}_\mu$ is not independent. In this case, there is no need to customize the settings of section I. So, the effective action is a functional of $A_\mu$ and $g_{ij}$. 

The above action also has nonrelativistic diffeomorphism invariance such that in addition to \eqref{transNR} for $g_{ij}$ and $A_\mu$, the transformation rules include the standard transformation for the non-Abelian gauge field
\begin{equation}\label{trans+a}
\delta {\A}_0=-\partial_k{\A}_0\xi^k- {\A}_k\dot\xi^k\,,\qquad
\delta {\A}_i=-\partial_k{\A}_i\xi^k-{\A}_k\partial_i\xi^k
\,.
\end{equation}
Here $\Psi$ transforms as $\delta\Psi=-\partial_k\Psi\xi^k$.  

A way to see that \eqref{trans+a} is standard is just to consider the variation of the non-Abelian gauge field under a relativistic infinitesimal transformation
\begin{equation*}
\delta {\cal A}_\mu=-\partial_\lambda {\cal A}_\mu\xi^\lambda-{\cal A}_\lambda\partial_\mu\xi^\lambda
\,,
\end{equation*}
with $\xi^\lambda=\bigl(0,\xi^k\bigr)$ and ${\cal A}_\mu=({\cal A}_0/c,{\cal A}_i)$. Note that in 
this case there are no subleading terms in the $1/c$ expansion.

This structure can easily be generalized to the case when the gauge group is $U(1)\times G$. We just assume that the 
transformation rules for the background fields are given by \eqref{transNR}, \eqref{trans+a} together with \eqref{transG} and \eqref{transGa}.

For $d=2$, the symmetry structure just described agrees nicely with that of the Pauli equation at $\text{g}=2$ (the value of the g-factor).\footnote{For a review, see \cite{FS,FW}.} In this case, 
$\Psi=\bigl(\begin{smallmatrix}\psi_+\\\psi_-\end{smallmatrix}\bigr)$ is the Pauli spinor and the $U(1)$ gauge field ${\cal A}_\mu$\footnote{We use $U(1)$ as isomorphic to $SO(2)$.} is given by 
\begin{equation}\label{GF}
	{\cal A}_0=-i\Bigl(\omega_0-\frac{1}{2m}B\Bigr)\sigma^3\,,\qquad
	{\cal A}_i=-i\omega_i\sigma^3\,,
\end{equation}
where $\sigma^3$ is the third Pauli matrix and the $\omega$'s are given by \eqref{SC}. It is easy to verify that such ${\cal A}_\mu$ obeys \eqref{trans+a}. The physical interpretation of the appearance of $B$ is clear, it is due to the Zeeman term. 

A couple of comments are in order. First, in the case at hand, one could define new fields
\begin{equation}\label{Apm}
A_0^\pm=A_0\pm{\cal A}^3_0\,,\qquad
A_i^\pm=A_i\pm{\cal A}^3_i\,,
\end{equation}
where ${\cal A}_0={\cal A}_0^3\sigma^3$ and ${\cal A}_i={\cal A}_i^3\sigma^3$. With this field redefinition, the Pauli equation reduces to two uncoupled Schr\"odinger equations. Clearly, under nonrelativistic diffemorphisms \eqref{xi} those fields transform as \eqref{transNR}. This is a simple example of how field redefinitions change the transformation rules. It is useful to keep this in mind when comparing different modifications of \eqref{transNR}.

Second, in the presence of the spin connection there is no need to modify the original transformation rules \eqref{transNR}. This implies that the constraints imposed by those transformation rules on the effective action remain unchanged compared to the original analysis of \cite{HS}, independently of the presence of the Zeemann interaction and independently of whether we assume the electrons to be spin-polarized or not (i.e. of whether we work with the $2$-component spinor in \eqref{Pauli} or with a single component field, assuming that a Hall fluid is completely spin-polarized due to the Zeemann effect). This is consistent with the discussion in \cite{WZ} and section VII of \cite{FS} where the structure of the effective theory is the same for spin-polarized and unpolarized Hall fluids, the only difference is in the concrete values of the parameters of the effective theory.

We conclude this section by giving a few simple examples of invariant actions. 

The first example is the Chern-Simons action
\begin{equation}\label{CS-nona}
	S=\text{Tr}\int dt d\mathbf{x} \,\varepsilon^{\mu\nu\lambda}\Bigl[
	{\cal A}_\mu\partial_\nu{\cal A}_\lambda+\frac{2}{3}{\cal A}_\mu{\cal A}_\nu{\cal A}_\lambda\Bigr]
	\,,
\end{equation}
with the trace in the fundamental representation of $G$. Obviously, \eqref{CS-nona} is invariant under the infinitesimal transformation \eqref{trans+a}.

If $G=U(1)$, we find two more examples
\begin{equation}\label{CS-a}
	S=\int dt d\mathbf{x} \,\varepsilon^{\mu\nu\lambda}{\cal A}_\mu\partial_\nu{\cal A}_\lambda
\,,
\end{equation}
which is trivially invariant, as well as
\begin{equation}\label{CS-a2}
S=\int dt d\mathbf{x} \,\Bigl[
	\varepsilon^{\mu\nu\lambda}\omega_\mu\partial_\nu{\cal A}_\lambda-\frac{\sqrt{g}}{2{\cal B}}g^{ij}\partial_i{\cal B}{\cal E}_j\Bigr]
	\,,
\end{equation}
where ${\cal E}_i=\dot{\cal A}_i-\partial_i{\cal A}_0$ and ${\cal B}=\text{e}^{ij}\partial_i{\cal A}_j$. What is particularly noteworthy about 
this invariant action is that it formally coincides with the two leading terms of the series \eqref{Son-1}.

Plugging ${\cal A}_0=\omega_0-\tfrac{1}{2m}B$, ${\cal A}_i=\omega_i$ into \eqref{CS-a2} and then using the invariance of \eqref{GSC}, we get
\begin{equation}\label{1/R}
	S=\int dt d\mathbf{x} \sqrt{g}\Bigl[RB-
	\frac{1}{R}g^{ij}\partial_iR\Bigl(2m(\dot{\omega}_j-\partial_j\omega_0)+\partial_jB
	\Bigr)\Bigr]\ ,
\end{equation}
where we used the relation ${\cal B}=R/2$. One can show that it is invariant under the nonrelativistic diffeomorphism transformations \eqref{transNR}. The Lagrangian in \eqref{1/R} can be regarded as a $1/R$ expansion. It would be interesting to see if it has any application to the Hall liquid living on a highly curved space. 

%__________________________________________________________ section 5
\section{The Continuity Equation and Hall Conductivity} 
\label{continuityequation}
\renewcommand{\theequation}{5.\arabic{equation}}
\setcounter{equation}{0}

Given symmetries of an action, it is straightforward to derive the corresponding conservation laws by Noether's theorem. Now let us consider the symmetries postulated in section \ref{intro}. 

Let $S$ be a functional of $A_\mu$ and $g_{ij}$ invariant under \eqref{transNR} and \eqref{transG}. First, by applying the gauge 
transformations \eqref{transG}, we get a current conservation law
\begin{equation}\label{current-cons}
	\partial_tJ^0+\oh\partial_t\ln g J^0+\nabla_i J^i=0\,,
\end{equation}
where $J^\mu=\frac{1}{\sqrt{g}}\delta S/\delta A_\mu$. Then, using the nonrelativistic diffeomorphism transformations \eqref{transNR}, in 2 dimensions we get\footnote{This was independently noticed in \cite{HMS}.} 
\begin{equation}\label{stress-cons}
m\partial_tJ_i+\oh m\partial_t\ln gJ_i+\nabla_j T^j_i=-J^0E_i+\text{e}_{ij}J^jB
\,,
\end{equation}
where $T^{ij}=\frac{2}{\sqrt{g}}\delta S/\delta g_{ij}$. 

If there is no more than one particle (quasiparticle) species contributing to the momentum flow, then the momentum density can be written in terms of the density of a conserved current
\begin{equation}\label{GWW}
T^{0i}=m J^i
	\,.
\end{equation}
With this identity, Eq.\eqref{stress-cons} turns into the continuity equation
\begin{equation}\label{cont-eq}
\partial_tT_i^0+\oh \partial_t\ln gT_i^0+\nabla_j T^j_i=\rho F_i
\,.
\end{equation}
Here $F_i$ is the Lorentz force $E_i-\text{e}_{ij}B v^j$, and $J^\mu=(-\rho,-\rho v^i)$. 

The above derivation of the continuity equation encourages the idea that the nonrelativistic diffeomorphism transformations \eqref{transNR} should be taken seriously.

It is interesting to ask wether the continuity equation can be derived from the energy-momentum conservation law
\begin{equation}\label{con-law}
\nabla_\mu T^{\mu\nu}=0
\,,
\end{equation}
where $T^{\mu\nu}=\tfrac{2}{\sqrt{g^{(3)}}}\delta S/\delta g_{\mu\nu}$. This is indeed the case. In fact, both the continuity equation and the current conservation law can be derived from \eqref{con-law}, if one identifies
\begin{equation}\label{T-exp}
T^{\mu\nu}=
\begin{pmatrix}
\,\,mcJ^0\,\,&{}\,\,&mJ^i\,\,\\
{}\,\,&{}\,\,&\,\,\\
\,\,mJ^i\,\,&{}\,\,& T^{ij}/c\,\,
\end{pmatrix}\,.
\end{equation}
In this derivation, one also has to use the ansatz \eqref{metricd+1} for the metric and take the $c \rightarrow \infty$ limit.

As usual, one can use equations \eqref{current-cons} and \eqref{stress-cons} to derive Ward identities for correlation functions. In particular, 
this allows one to derive a relation between the stress response tensor and the conductivity tensor \cite{Read}. For small frequencies and wave numbers, this results in a relation between the Hall viscosity and the $q^2$ part of the Hall conductivity \cite{HS}. More precisely, according to \cite{Read}, equation \eqref{stress-cons} in flat space and in the absence of an electric field, i.e.
\begin{equation}\label{continuity}
	\bigl(m\delta_{ij}\partial_t-B\varepsilon_{ij}\bigr)J_j+\partial_j T_{ij}=0\ ,
\end{equation}
leads to the following result for the $q^2$ part of the conductivity tensor as $\omega\rightarrow 0$
\begin{equation}\label{condactivity}
	\sigma_{ij}^{(2)}(q)\sim \frac{1}{B^2}\left[
	\Bigl(\zeta+i\frac{B}{m} \omega^{-1} \kappa_{\text{in}}\Bigr)\varepsilon_{in}\varepsilon_{jk}q^nq^k+
	\eta^{\text{sh}}q^2\delta_{ij}+
	\Bigl(\eta^{\text{H}}-\kappa_{\text{in}}+2i\frac{m}{B}\omega\eta^{\text{sh}}\Bigr)q^2\varepsilon_{ij} \right]
\,.
\end{equation}
Here $\zeta,\,\eta^{\text{sh}},\,\eta^{\text{H}}$ are the bulk, shear and Hall viscosities, respectively. $\kappa_{\text{in}}$ represents a 
contribution of the inverse internal compressibility. The final (antisymmetric) term proportional to $q^2\varepsilon_{ij}$ (for the case of an $ \eta^{\text{sh}}$ which does not diverge for $\omega \rightarrow 0$ so that the last term $\sim \omega$ can be neglected) was derived in \cite{HS} from the effective action 
\begin{equation}\label{SH-action}
S_{\text{eff}}=\frac{\kappa}{2\pi} S_1-S_2\ ,
\end{equation}
with $S_1$ and $S_2$ given by \eqref{Son-1}. In terms of this action one has $\eta^{\text{H}}=\frac{\kappa B}{4\pi}$ and $\kappa_{\text{in}}=mB\epsilon''(B)$. 

We end this section with the following observation. Consider an effective action which differs from \eqref{SH-action} by a term proportional to \eqref{EH3b}
\begin{equation}
S_{\text{eff}}\rightarrow S_{\text{eff}}+\Delta S_3\,.
\end{equation}
Here $S_3$ is given by \eqref{EH3b} and $\Delta$ is a constant. At order $\epsilon^2$ in the expansion of \cite{HS}, this amounts to simultaneously shifting $\kappa \rightarrow \kappa + 2 \pi \Delta$ and $\epsilon(B) \rightarrow \epsilon(B) + \frac{\Delta}{4 m} B^2$. As a result, we get
\begin{equation}\label{mod}
\eta^{\text{H}}\rightarrow\eta^{\text{H}}+\oh \Delta B
\,,\qquad \kappa_{\text{in}}\rightarrow\kappa_{\text{in}}+\oh\Delta B
\,.
\end{equation}
Comparing with the expression \eqref{condactivity} for the conductivity tensor, we note that two effective theories whose actions differ by the addition of a multiple of \eqref{EH3b} have the same coefficient in front of $\varepsilon_{ij}q^2$, while those in front of $\varepsilon_{in}\varepsilon_{jk} q^nq^k$ are different.

%__________________________________________________________ section 6
\section{Ho\v{r}ava-Lifshitz Gravity with $U(1)$ gauge Symmetry} 
\renewcommand{\theequation}{6.\arabic{equation}}
\setcounter{equation}{0}
\label{sec_horava_lifshitz}

The above issues are similar to those in a theory of gravity with anisotropic scaling called Ho\v{r}ava-Lifshitz gravity. Its primary goal is to describe the dynamics of the spatial metric $g_{ij}(t,\mathbf{x})$, in a way which is invariant under foliation-preserving diffeomorphisms
\begin{equation}\label{foll-diffs}
\delta t=f(t)\,,\qquad
\delta x^i=\xi^i(t,\mathbf{x})
\,.
\end{equation}
In this case, the spacetime metric $g_{\mu\nu}$ is written in the Hamiltonian form 
\begin{equation}\label{ADM}
	ds^2=-N^2 c^2dt^2+g_{ij}\bigl(dx^i+N^icdt\bigr)\bigl(dx^j+N^j cdt\bigr)
	\,,
\end{equation}
where $N$ is a lapse function and $N^i$ is a shift vector.\footnote{Note that we use different coordinates in \eqref{ADM} and \eqref{adm}, which is why we use a different notation for the shift vector here.} It is important to remember that in the projectable version of Ho\v{r}ava-Lifshitz gravity, 
which we will discuss here, $N$ is only a function of time. 

At leading order in time derivatives, the effective action is given by \cite{Horava}
\begin{equation}\label{HLG}
	S=\frac{1}{\kappa^2}\int dtd{\mathbf{x}}\sqrt{g} N
	\Bigl[K_{ij}K^{ij}-\lambda K^2+\beta\bigl(R-2\Lambda \bigr)
	\Bigr]
	\,,
\end{equation}
where $K_{ij}=\oh N^{-1}\bigl(\partial_0g_{ij}-\nabla_i N_j-\nabla_jN_i\bigr)$, $K=g^{ij}K_{ij}$, and $\lambda,\,\beta$ are couplings.

With the metric expansion \eqref{metricd+1}, the lapse function and shift vector take the form
\begin{equation}\label{ADMa}
N=\sqrt{1-\frac{2A_0}{mc^2}+\frac{A_i A^i}{m^2c^2}}\,,\qquad
N_i=\frac{A_i}{mc}
	\,.
\end{equation}
Once we specified the lapse function in terms of the gauge fields, we need to make a choice of gauge which is consistent with the requirement 
that $N$ is independent of $\mathbf{x}$. We choose 
\begin{equation}\label{gauge}
	2m\partial_i A_0-\partial_i(A_jA^j)=0\,.
\end{equation}
Such a choice doesn't fix the gauge uniquely. There is a remaining gauge transformation 
\begin{equation}\label{gauge-trans}
	\delta A_0=-\dot\alpha\,,\qquad \delta A_i=0\,,
\end{equation}
with $\alpha$ a function of time.

For $d=2$, just as in our study of the Einstein-Hilbert action in section 2, we can take the limit of the functional \eqref{HLG} as $c\rightarrow\infty$. In this case, we regard $\kappa^{-2}$ as $mc^2$, and similarly we subtract the leading term proportional to the Euler characteristic $\chi$. The action is 
then (setting $\beta=1$ and $\Lambda=0$)
\begin{equation}\label{HLGa}
S=m\int dt d\mathbf{x} \sqrt{g}\Bigl[-\frac{1}{4}\dot{g}^{ij}\dot{g}_{ij}-\frac{\lambda}{4}(g^{ij}\dot{g}_{ij})^2-
	\frac{1}{m}\Bigl(\lambda A^i\partial_t\nabla_i\ln g-\dot{g}^{ij}\nabla_i A_j\Bigr)+
	\frac{1}{m^2}\Bigl((1-\lambda)(g^{ij}\nabla_iA_j)^2+\frac{1}{4}F_{ij}F^{ij}-\oh A_iA^i R\Bigr)\Bigr]
	\,.	
\end{equation}
Evidently, with the relation \eqref{EH3a}, the $U(1)$ gauge symmetry is restored if the value of $\lambda$ is set to $1$ and a term 
$\tfrac{R}{m}(\tfrac{1}{2m}A_iA^i-A_0)$ is added to the original Lagrangian density.\footnote{A similar conclusion was reached through the Noether procedure in \cite{Horava}.} If so, then the resulting action takes the form \eqref{EH3b} which is manifestly gauge-invariant. 

Returning to our main theme, an alternative way to find actions invariant under the transformation rules \eqref{transNR} and \eqref{transG} is to first define the fields (up to an additive constant for ${\cal N}$)
\begin{equation}\label{fields}
{\cal N}=A_0-\frac{1}{2m}A_iA^i\,,\qquad
{\cal K}_{ij}=\dot g_{ij}-\frac{1}{m}\bigl(\nabla_i A_j+\nabla_j A_i\bigr)\,.
\end{equation}
This is motivated by the fact that under nonrelativistic diffeomorphisms \eqref{transNR}, ${\cal N}$ and ${\cal K}_{ij}$ are a scalar and a tensor, respectively. Indeed, it is easy to see that
\begin{equation}\label{non-rel-trans}
	\delta {\cal N}=-\partial_k{\cal N}\xi^k\,,\qquad \delta {\cal K}_{ij}=-\partial_k {\cal K}_{ij}\xi^k-{\cal K}_{kj}\partial_i\xi^k-{\cal K}_{ik}\partial_j\xi^k
	\,.
\end{equation}

To construct another scalar, we consider a field of the form ${\cal N}'=A_0'-\frac{1}{2m}A_i'A'{}^i$, where $A'$ is given by \eqref{newA}. This ensures that ${\cal N}'$ is a scalar. After expressing $A'$ in terms of $A$, we find that since $A_0-\frac{1}{2m}A_iA^i$ and $\omega_i\omega^i$ are scalars, 
\begin{equation}\label{newscalar}
	{\cal M}=\omega_0-\frac{1}{m}\omega_i A^i-\frac{1}{2m}B
	\,
\end{equation}
must be a scalar too.

For given ${\cal N}$, ${\cal M}$, ${\cal K}_{ij}$, $\omega_i$, and $g_{ij}$, we can construct an action invariant under nonrelativistic diffeomorphisms. In general, the action might not be invariant under the gauge transformations \eqref{transG} as well as the local $SO(2)$ transformations \eqref{Loretz}, and further requiring invariance constrains the number of independent terms in the action.\footnote{Note that invariance under nonrelativistic diffeomorphisms plays the same role in section 2.}

Our remaining goal will be to develop this approach at the two derivative level. There are five independent invariants of dimension $m^sc^4$, where $s$ is an arbitrary integer, ${\cal K}^2,\,{\cal K}_{ij}{\cal K}^{ij},\, {\cal N}R$, ${\cal N}g^{ij}\nabla_i\omega_j$, and ${\cal M}^2$. So the action invariant under nonrelativistic diffeomorphisms takes the form
\begin{equation}\label{inv-action}
	S=\int dtd\mathbf{x}\sqrt{g}
	\Bigl[a_1{\cal K}^2+a_2{\cal K}_{ij}{\cal K}^{ij}+a_3\,{\cal N}R+a_4\,{\cal N}g^{ij}\nabla_i\omega_j+
	a_5{\cal M}^2
	\Bigr]
	\,,
\end{equation}
where $a_i$ are couplings. 

First, requiring $S$ to be invariant under the local $SO(2)$ transformations, we get 
\begin{equation}\label{a}
	a_4=a_5=0
	\,.
\end{equation}
Then, requiring invariance under the $U(1)$ gauge transformations, we learn that the $a_i$'s must obey 
\begin{equation}\label{relations}
a_2=-a_1\,,\qquad
a_3=-\frac{4}{m}a_1\,.
\end{equation}
In this case a short calculation shows that the action \eqref{inv-action} reduces to \eqref{EH3b}, as expected.

An important point is that as long as $S$ is also a functional of $\omega_0$ and $\omega_i$, requiring invariance under the local $SO(2)$ transformations imposes additional constraints on the action. If instead $S$ is a functional of the metric and gauge field only, then just as in the above example from Ho\v{r}ava-Lifshitz gravity, one can restrict its form by requiring gauge invariance. 

%___________________________________________________________ section 7
\section{Concluding Comments} 
\renewcommand{\theequation}{7.\arabic{equation}}
\setcounter{equation}{0}

\noindent (i) Recently, new transformation rules under nonrelativistic diffeomorphisms \eqref{xi} have been proposed \cite{Son}
\begin{equation}\label{trans-new}
\delta A_0=-\partial_k A_0\xi^k- A_k\dot\xi^k+\frac{\text{g}}{4}\text{e}^{ij}g_{jk}\nabla_i\dot{\xi}^k
\,,\quad\delta A_i=-\partial_k A_i\xi^k- A_k\partial_i\xi^k-mg_{ik}\dot\xi^k\,,\quad
\delta g_{ij}=-\partial_k g_{ij}\xi^k-g_{kj}\partial_i\xi^k-g_{ik}\partial_j\xi^k\,.
\end{equation}
In fact, this is a one-parameter deformation of \eqref{transNR} whose parameter is $\text{g}$.

Now let us discuss alternative ways to address the issue of deformation. For the sake of simplicity, we consider only the case of two spatial dimensions.

As we have already remarked in section IV, from the point of view of fermions coupled to gravity, there is no need to modify the original transformation rules. Introducing the field ${\cal A}_\mu$ representing the spin connection allows us to do so (which is explicitly done for ${\text g}=2$ in \eqref{GF}). 

Another way to avoid modifying the original transformation rules \eqref{transNR} is to make the following field redefinition
\begin{equation}\label{new-A}
 A'_0=A_0+\oh\text{g}\,\omega_0\,,\qquad
A'_i=A_i+\oh\text{g}\,\omega_i
\,,
\end{equation}
where $\omega_0$ and $\omega_i$ are given by \eqref{SC}. With the aid of \eqref{spinconNR}, we find that $A'_0$ and $A'_i$ transform according to the original transformation rules \eqref{transNR}.

Note that demanding that $A'_\mu$ defined in \eqref{new-A} does not transform under local Lorentz transformations would amount to a twist that leads to new transformation rules for the dynamical fields under local Lorentz transformations. For example, the upper component of the Pauli spinor would then transform as $\delta\psi_+=-i(1+\oh\text{g})\Lambda\psi_+$. The notion of twisting was intensively discussed in the context of topological and supersymmetric field theories, but this is not what we are interested in here.\footnote{We thank D. Plencner and I. Sachs for a discussion of this issue.}

\noindent (ii) To leading order in $c$, when taking the nonrelativistic limit of a covariant theory, the resulting theory might capture some of 
the physics and become more amenable to mathematical treatment. However, it loses much of its ability to make predictions. For example, in flat space 
the action corresponding to the Pauli equation reduces to 
\begin{equation}\label{Pauli-reduced}
S=\int dtd\mathbf{x}\sqrt{g}\biggl[\frac{i}{2}\Psi^\dagger\overset\leftrightarrow{\partial}_t\Psi+
\Psi^\dagger(A_0-\frac{1}{2m}\vec{B}\cdot\vec{\sigma})\Psi-
\frac{1}{2m}
\bigl(\partial_i\Psi^\dagger+i\Psi^\dagger A_i\bigr)
\bigl(\partial_i\Psi-iA_i\Psi\bigr)
\biggr]
\,.
\end{equation}
So, one has just the spinors coupled to an electromagnetic field and the term describing the Zeeman splitting but all others, including the term describing spin-orbit interactions, are gone. This problem calls for the computation of $1/c$ corrections to the transformations \eqref{transNR}.

\noindent (iii) In section IV, the treatment of the case with extended gauge symmetry is motivated by coupling spinors to gravity. In that case the group of local Lorentz 
transformations appears. Alternatively, one can consider a relativistic action which is invariant under gauge transformations (Abelian or non-Abelian) and then take the nonrelativistic limit \cite{JK}.

\noindent (iv) Certainly, there are terms in the effective action which depend only on the gravity fields $V^i$ and $g_{ij}$. At the two derivative level, these terms can be read off from \eqref{inv-action} with ${\cal N}=1$ and ${\cal K}_{ij}=\dot g_{ij}+\nabla_i V_j+\nabla_j V_i$.

%__________________________________________________________________
\begin{acknowledgments}

This work was supported in part by the DFG grants HA 3448/6-1 and HA 3448/3-1. We would like to thank I.S.\ Burmistrov, D.\ Plencner, I.\ Sachs, D.T.\ Son and G.E.\ Volovik for useful discussions. O.A.\ thanks the Arnold Sommerfeld Center for Theoretical Physics at LMU for its hospitality during 
much of this work.
\end{acknowledgments}
%__________________________________________________________________
\appendix
\section{Notation, Conventions, and  Useful Formulas}
\label{notation}
\renewcommand{\theequation}{A.\arabic{equation}}
\setcounter{equation}{0}

\subsection{Notation and Conventions}

Throughout this paper we use the following notations and conventions: Latin indices $i,\,j,\,k$, and so on generally run over $d$ spatial coordinate labels. Repeated indices are summed unless otherwise indicated. A dot over any quantity denotes a time derivative of that quantity.

For $d=2$, Greek indices $\mu ,\,\nu ,\,\lambda$, and so on run over the three space-time coordinate labels in a general coordinate system such that $x^0=ct$. Latin indices $a,\,b,\,c$, and so on run over the two spatial inertial coordinate labels $1,\,2$. The metric in an inertial coordinate system is $\text{diag}(1,\,1)$.

The reduced Planck constant and the electric charge (gauge coupling constant) are taken to be unity.

We use the following conventions for covariant derivatives 
\begin{equation}\label{derivatives}
	\nabla_\mu A^\nu=\partial_\mu A^\nu+\Gamma^\nu_{\mu\lambda}A^\lambda\,,\qquad
	\nabla_\mu A_\nu=\partial_\mu A_\nu-\Gamma^\lambda_{\mu\nu}A_\lambda
	\,,
\end{equation}
where the Levi-Civit\`{a} connection is $\Gamma^\lambda_{\mu\nu}=\oh g^{\lambda\rho}\bigl(\partial_\mu g_{\nu\rho}+\partial_\nu g_{\mu\rho}-
\partial_\rho g_{\mu\nu}\bigr)$.

The Ricci and scalar curvature tensors are given by 
\begin{equation}\label{Ricci}
	R_{\mu\nu}=\partial_\lambda\Gamma^\lambda_{\mu\nu}-\partial_\mu\Gamma^{\lambda}_{\nu\lambda}+\Gamma^\lambda_{\mu\nu}\Gamma^\rho_{\lambda\rho}-
	\Gamma^\lambda_{\mu\rho}\Gamma^\rho_{\nu\lambda}
	\,,\qquad
	R=g^{\mu\nu}R_{\mu\nu}\,.
\end{equation}

For $d=2$, we use the following conventions for the completely antisymmetric symbols $\varepsilon^{12}=1$ and $\varepsilon^{012}=1$. It is also useful to 
have the completely antisymmetric tensor $\text{e}^{ij}=\varepsilon^{ij}/\sqrt{g}$, with $g=\det g_{ij}$.

\subsection{Some Useful Formulas}

Up to $1/c^3$ terms, the $1/c$ expansion of the inverse metric of \eqref{metricd+1} is given by 
\begin{equation}\label{Metric2+1}
g^{\mu\nu}=
\begin{pmatrix}
-1-{\displaystyle\frac{2A_0}{mc^2}+\frac{A_nA^n}{m^2c^2}}\,\,&{}\,\,&{\displaystyle\frac{A^i}{mc}}\,\,\\
{}\,\,&{}\,\,&\,\,\\
{\displaystyle\frac{A^i}{mc}}\,\,&{}\,\,& g^{ij}-{\displaystyle\frac{A^iA^j}{m^2c^2}}\,\,
\end{pmatrix}\,.
\end{equation}

In this case, the expansion of the three-dimensional Ricci tensor $R^{(3)}_{\mu\nu}$ is 
\begin{equation}\label{Ricci3a}
	\begin{split}
R^{(3)}_{00}=&-\frac{1}{2c^2}g^{ij}\ddot{g}_{ij}-\frac{1}{4c^2}\dot{g}^{ij}\dot{g}_{ij}+\frac{1}{mc^2}\nabla_i E^i+\frac{1}{4m^2c^2}F_{ij}F^{ij}\,,
\qquad
R^{(3)}_{0i}=-\frac{1}{2c}\partial_t\partial_i\ln g+\frac{1}{2c}g^{nm}\nabla_n\bigl(\dot{g}_{im}+\frac{1}{m}F_{im}\bigr)\,\\
R^{(3)}_{ij}=&R_{ij}+
\frac{1}{4c^2}g^{nm}\Bigl(\partial_t(g_{nm}\dot{g}_{ij})-2\dot{g}_{in}\dot{g}_{jm}\Bigr) 
-\frac{1}{mc^2}\Bigl(\frac{1}{\sqrt{g}}\partial_t\bigl(\sqrt{g}\nabla_{(i}A_{j)}\bigr)-\nabla_i\nabla_jA_0+\oh\nabla_n(\dot{g}_{ij}A^n)
+\dot{g}^{nm}\nabla_m g_{n(i}A_{j)}\Bigr)\\
+&\frac{1}{2m^2c^2}
\Bigl(2\nabla_n(A^n\nabla_{(i}A_{j)})-\nabla_i\nabla_j (A_nA^n)+
g^{nm}\bigl(\nabla_i A_n\nabla_j A_m-\nabla_nA_i\nabla_mA_j\bigr)
\Bigr)\,,
\end{split}
\end{equation}
where $M_{(ij)}=(M_{ij}+M_{ji})/2$ and $F_{ij}=\partial_iA_j-\partial_jA_i$.

The electric and magnetic fields are expressed in terms of the components of $A_\mu$
\begin{equation}\label{EB}
	E_i=\dot{A}_i-\partial_i A_0\,,\qquad B=\text{e}^{ij}\partial_i A_j\,.
\end{equation}
It is also useful to have the following formulas at hand for their transformations under nonrelativistic diffeomorphisms:
\begin{equation}\label{B}
	\delta B= -\partial_k B\xi^k-m\text{e}^{ij}g_{jk}\nabla_i\dot{\xi}^k\,,\qquad
	\delta E_i=-\partial_kE_i\xi^k-E_k\partial_i\xi^k+\text{e}_{ik}B\dot{\xi}^k-m\partial_t(g_{ik}\dot{\xi}^k)\,.
\end{equation}
Therefore $B$ and $E_i$, being gauge invariant, don't transform as a scalar and a vector under the nonrelativistic diffeomorphism transformations \eqref{transNR}.

A zweibein $e^a_i$ is invariant under the $U(1)$ gauge transformations, while under the nonrelativistic diffeomorphisms it transforms as a vector
\begin{equation}\label{zweibein}
\delta e^a_i=-\partial_k e^a_i\xi^k-e_k^a\partial_i\xi^k
\,.
\end{equation}
Hence the spin connections defined in \eqref{SC} transforms as
\begin{equation}\label{spinconNR}
	\delta\omega_0=-\partial_k\omega_0\xi^k-\omega_k\dot{\xi}^k-\oh\text{e}^{ij}g_{jk}\nabla_i\dot{\xi}^k
	\,,\qquad
	\delta\omega_i=-\partial_k\omega_i\xi^k-\omega_k\partial_i\xi^k
	\,.
\end{equation}

The scalar curvature is expresses in terms of the spin connection
\begin{equation}\label{R}
R=2\,\text{e}^{ij}\partial_i\omega_j
\,.
\end{equation}
In addition, the Euler characteristic of a two-dimensional Riemann surface is given by 
$\chi=\frac{1}{2\pi}\int d^2x\,\varepsilon^{ij}\partial_i\omega_j$.

It is also useful to record the transformation rules under local $SO(2)$ rotations. In doing so, we must remember to include time 
dependence. The rules thus read
\begin{equation}\label{Loretz}
	\delta\omega_0=-\dot{\Lambda}\,,
	\qquad
	\delta\omega_i=-\partial_i\Lambda
	\,,
\end{equation}
with $\delta e_i^a=\Lambda\varepsilon^{ab}e^b_i$. 

One way to prove that the action \eqref{EH3b} is invariant under the nonrelativistic diffeomorphism transformations is to first rewrite it (up to a constant factor) as 
\begin{equation}\label{EH3d}
	S=m\int dt d^2x \sqrt{g}
	\biggl[e^i_a\ddot{e}^a_i+\frac{2}{m}\text{e}^{ij}\omega_i E_j+
	2\Bigl(\omega_0-\frac{1}{2m}B\Bigr)^2
	\biggr]\,
\end{equation}
and then to use	
\begin{equation}\label{Omegab}
\delta\Bigl(\omega_0-\frac{1}{2m}B\Bigr)=
-\partial_k\Bigl(\omega_0-\frac{1}{2m}B\Bigr)\xi^k-\omega_k\dot{\xi}^k
\,.
\end{equation}
The fact that this combination of $\omega_0$ and $B$ transforms like a 1-form under nonrelativistic diffeomorphisms was also noticed in \cite{HMS}.
%___________________________________________________________________

\end{document}